\documentclass[prb,english,twocolumn,amsmath,showpacs]{revtex4}
\usepackage[latin1]{inputenc}
\usepackage{graphicx}% Include figure files
\usepackage{babel}

\begin{document}

\title{Evidence of {\sl s\/}-Wave Subdominant Order Parameter in YBCO from Break Junction Tunneling Spectra}

\author{A. I.\ Akimenko,$^{1,2,}$\cite{AkinekoAI} F.\ Bobba,$^{1}$ F.\ Giubileo,$^{1}$ V. A.\ Gudimenko,$^{2}$ A.\ Scarfato,$^{1}$ and A. M.\ Cucolo$^{1}$}

\affiliation{$^{1}$Department of Physics and INFM Unit, University of Salerno, Via S. Allende; 84081 Baronissi, Italy \\
$^{2}$B.\ Verkin Institute for Low Temperature Physics and Engineering
National Academy of Sciences of Ukraine, 47 Lenin Ave., 61103, Kharkiv,
Ukraine}

\date{\today{}}

\begin{abstract}
The tunneling spectra of YBa$_{2}$Cu$_{3}$O$_{7-\delta}$ break-junctions have been investigated for the tunneling direction close to the node one. The zero-bias conductance peak (ZBCP) and Josephson current have been studied with temperature and magnetic field. The observed deep splitting of ZBCP which starts at $T _S$$\rm <$20--30\,K is in agreement with the theory for the {\it d$_{x^{2}-y^{2}}$$\pm$is} order parameter [Y. Tanuma, Y. Tanaka, and S. Kashiwaya, Phys. Rev. B \textbf{64}, 214519 (2001)]. The low (0.04\,T) magnetic field depresses significantly such splitting. The $1/T$ temperature dependence of maximum Josephson current goes to saturation at $T$$<$$T_S$ also confirming the mixed order parameter formation.


\end{abstract}
\maketitle

For the $d$-wave superconductors, theory predicts specific quasiparticle bound states (Andreev bound states) near scattering structures such as surfaces, interfaces and other defects \cite{Kashiwaya}. In these areas an order parameter (OP) may change significantly and subdominant OP may appear leading to a mixed OP (such as {\it d$_{x^{2}-y^{2}}$$\pm$is} or {\it d$_{x^{2}-y^{2}}$$\pm$id$_{xy}$} with the spontaneous breaking of time reversal symmetry (BTRS) \cite{Sigrist,Matsumoto,Zhu99,Tanuma}.

Andreev bound states manifest themselves in different tunneling spectra as a zero bias conductance peak in agreement with the theory for the {\it d$_{x^{2}-y^{2}}$\/}-wave pairing \cite{Kashiwaya}. In the case of breaking of time reversal symmetry (due to magnetic field or subdominant OP), splitting of ZBCP was predicted \cite{Tanuma,Fogelstrom97,Zhu98} and also observed in several experiments \cite{Covington,Aprili,Greene,Krupke,Giubileo00}. Theory shows the different kind of splitting for the $is$ and {\it id$_{xy}$} subdominant OP \cite{Tanuma,Rainer}. This question has not been studied in experiments up to now.

The Josephson critical current may also give information about subdominant OP presence. Its temperature dependence is predicted to saturate at temperatures $T$$<$$T_S$ ($T_S$ is the critical temperature for subdominant OP) \cite{Tanaka}.

It is interesting to note that some theories predict splitting of ZBCP without any subdominant OP \cite{Fogelstrom98} and even BTRS \cite{Morr}.

To solve the problem we have investigated the \mbox{$S$--$I$--$S$} Josephson junctions. The break-junction method for a thin film was applied and the tunneling spectra with the deep splitting of ZBCP at temperatures up to 20--30\,K have been registered. The analysis of the temperature and magnetic field dependences says in favor of the $is$ subdominant order parameter presence. The maximum Josephson current also saturates at $T$$<$$T_S$.

The tunnel junctions were produced by applying the special break-junction technique \cite{Akimenko} to highly biepitaxial $c$-axis oriented YBa$_{2}$Cu$_{3}$O$_{7-\delta}$ thin films (thickness $\rm \approx$200\,nm), d.c. sputtered on (001) SrTiO$_{3}$ substrates \cite{Beneduce}. Electrical characterization showed critical temperatures $T_C$ $(\rho=0)$ $>$ 91\,K and $\Delta$$T_C$$<$1\,K. To determine the lateral lattice alignment between the films and the substrates the $X$-ray pole figure analysis was used \cite{Cucolo}. The stripe-like samples (with the [110]-direction long side) were glued to a metallic bending plate by the epoxy glue. A special epoxy cover over the whole sample was applied to make the construction stable with the time and temperature change (more details see in Refs.\,\onlinecite{Akimenko}\,and\,\onlinecite{Cucolo}). As a result we were able to investigate a single break junction in about a weak time, with the only small change in its resistance in the temperature range 4.2--120\,K \cite{Giubileo01}. To maximize the tunnel current along the node direction a straight groove was scratched in the central part of the covered sample (perpendicular to [110] direction), where the bending is maximum. By bending with a differential screw at low temperature, it is possible to crack the substrate together with the film along the groove and smoothly adjust the junction resistance. The optical microscope study showed that the fracture direction can deviate from the straight line only about 10°. More experimental details can be found in Ref.\,\onlinecite{Cucolo}.

\begin{figure}
\begin{center}
\includegraphics[width=9.5cm,angle=0]{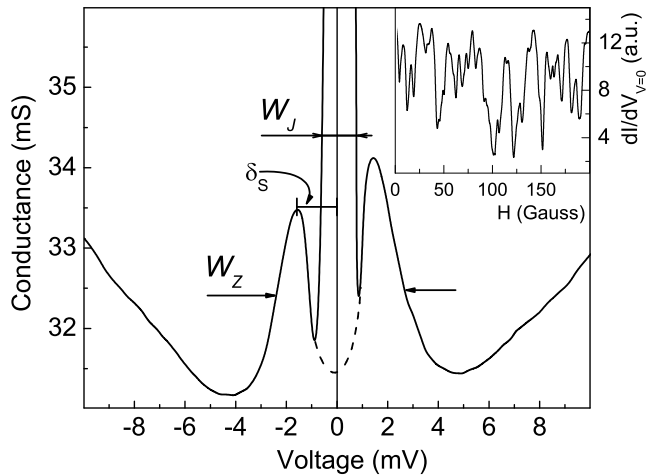}
\end{center}
\caption{Tunneling spectrum ($dI/dV$ vs $V$) of YBCO break-junction at $T$=10\,K (solid line). The dashed line is a possible minimum around $V$=0 drawn by hand. The Josephson peak (with the width {\sl W$_J$}$\approx$$\pm$1\,mV) superimposes on the double peak structure ({\sl W$_Z$}$\approx$$\pm$2.5\,mV). The width here is only qualitative characteristic of the peak structure. $\delta$$_S$ shows a position (from $V$=0) of the peak in the double peak structure. To understand the relative intensity of the peaks see Fig.\,\ref{Ttunneling}.\\
Inset:	Magnetic field dependence of the zero-bias conductance. The external field was applied parallel to the $c$\,-axis direction.}
\label{TunnelingSpectra}
\end{figure}

In Fig.\,\ref{TunnelingSpectra} we show the low-bias tunneling spectrum $dI/dV$ vs $V$ of the YBCO break junction at $T$=10\,K measured by standard modulation technique. One can observe the simultaneous presence of two peak structures. Indeed, a well developed, narrow peak (with the width {\sl W$_J$}$\approx$$\pm$1\,mV) centered at zero energy appears superimposed to a less pronounced, wider double-peak structure {\sl W$_Z$}$\approx$$\pm$2.5\,mV). In addition to these the wide gap-related maxima (or the bound states with nonzero energy \cite{Kashiwaya}) around $\pm$15\,mV are observed that shifts towards lower biases for increasing temperature and disappears at $T$$\rightarrow$$T_C$ \cite{Giubileo01}. The similar peak structure with ZBCP (without splitting) and gap-related peak was also found in Ref.\,\onlinecite{Iguchi} for the close to [110] direction tunneling in the \mbox{$N$--$I$--$S$} ramp-edge junctions.

The narrow peak centered at $V$=0 is mostly due to the Josephson direct current though it corresponds to a smeared current step at $V$=0. The more the junction resistance the less its relative intensity. However, the most decisive argument in favor of the Josephson tunneling is the magnetic field dependence of the conductance at $V$=0 shown in inset of Fig.\,\ref{TunnelingSpectra}. The similar oscillating behavior was also found for the Josephson critical current in junctions with the nonuniform current-density distribution \cite{Hilgenkamp}. In our case, the nonuniform current may be also due to some deviations from the planar configuration of the junction. However, the similar current step with the finite conductance at $V$=0 was earlier observed in the YBCO grain boundary junctions \cite{Medici} as well as in the YBCO and Nb break-junctions \cite{Cucolo,Giubileo01,Iguchi,Hilgenkamp,Medici,Muller}. Around junction $T_C$, thermal and external fluctuations can induce the nonzero resistance since the Josephson coupling energy $E_J$=$hI_C/2e$ is comparable with the thermal energy $k_B$$T$. However, at least for our low resistance junctions ($R_N$=20--100\,$\rm\Omega$) at liquid helium temperature $E_J$ was greater than $k_B$$T$ by a factor of 20. On the other hand, as far as we know, the fluctuation effect on Josephson current for the $d$-wave superconductors is not studied jet.

\begin{figure}
%\begin{center}
\includegraphics[width=9.5cm,angle=0]{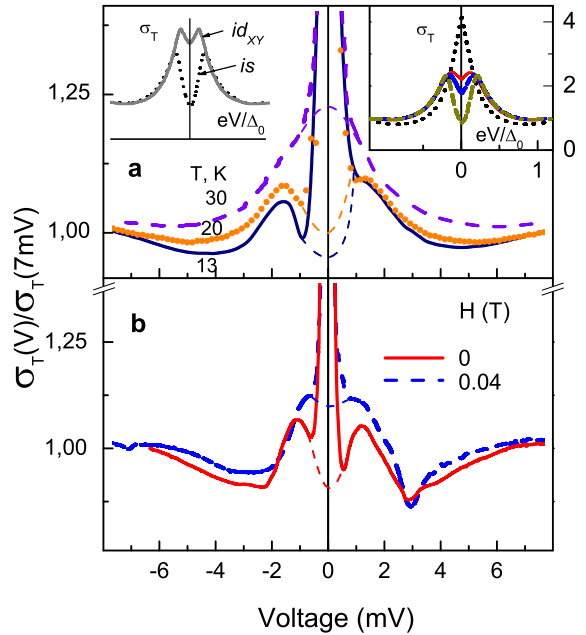}
%\end{center}
\caption{Temperature (a) and magnetic field (b) dependences of normalized tunnel conductance {\large$\sigma$}$_T$$(V)$ at low temperatures $T$ and magnetic field $H$. Normalizations of {\large$\sigma$}$_T$=$dI/dV$ are done for $T$=13\,K(a) and $H$=0(b) at $V$=7\,mV. The thin dashed line is a possible form of the spectrum under the Josephson peak.\\
Panel: Comparison of calculated tunneling conductance {\large$\sigma$}$_T$$(eV/\Delta_0)$ of \mbox{$N$--$I$--$S$} junction for the node direction tunneling and for the {\it d$_{x^{2}-y^{2}}$$\pm$is} and {\it d$_{x^{2}-y^{2}}$$\pm$id$_{xy}$} order parameters \cite{Tanuma}. Bath temperature $T$$/$$T_C$=0.05. $T_S$$/$$T_C$=0.2.\\
Inset: Calculated \mbox{$N$--$I$--$S$} junction tunneling conductance {\large$\sigma$}$_T$$(eV/\Delta_0)$ for the {\it d$_{x^{2}-y^{2}}$$\pm$is\/} state for the node direction tunneling \cite{Tanuma}. $T$$/$$T_C$=0.05, 0.10, 0.12, 0.13 starting from bottom at $eV$=0. $T_S$$/$$T_C$=0.2.}
\label{Andreev}
\end{figure}

\begin{figure}
%\begin{multicol}{1}
%\begin{center}
\includegraphics[width=9cm,angle=0]{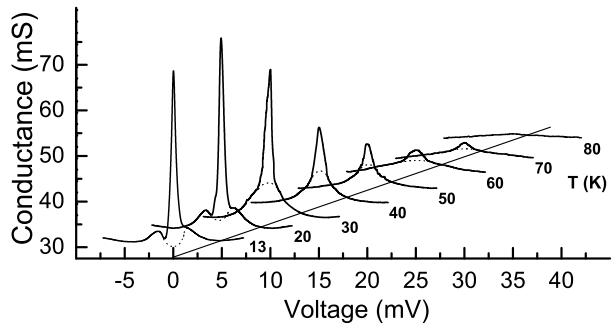}
%\end{center}
\caption{Temperature dependence of tunneling spectrum $dI/dV$ vs $V$. The curves at $T$$>$13\,K have been successively shifted along the bias (with a 5\,mV step) and conductance axes (along the thin solid line). The dashed line is a possible background under Josephson peak.}
\label{Ttunneling}
%\end{multicol}
\end{figure}

The double-peak structure with the width {\sl W$_Z$} looks like the expected Andreev bound states structure for the case of subdominant OP presence \cite{Kashiwaya}. The possible form of the structure under the Josephson peak is shown by dashed line. In Figs.\,\ref{Andreev}\,and\,\ref{Ttunneling} we show change of the similar structures with temperature and low magnetic field. One can see that the structure, observed at low temperatures, disappears with the temperature raise at $T$ between 20 and 30\,K transforming into the single wide peak (Fig.\,\ref{Andreev}a). Such splitting of ZBCP with the deep minimum is only predicted for the {\it d$_{x^{2}-y^{2}}$$\pm$is} order parameter (panel of Fig.\,\ref{Andreev}a) \cite{Tanuma,Rainer}. The experimental temperature dependence is similar to the calculated one in Ref.\,\onlinecite{Tanuma} (inset in Fig.\,\ref{Andreev}a). We should note that for the \mbox{$S$--$I$--$S$} junctions investigated here, the relative intensity of extremums must be more than that for the calculated in Ref.\,\onlinecite{Tanuma} \mbox{$N$--$I$--$S$} junctions \cite{Wolf}. Nevertheless, the alternative {\it d$_{x^{2}-y^{2}}$$\pm$id$_{xy}$}  order parameter will not give so deep minimum that are found in our experiments.

Thus, for the first time we have clear evidence in favor of the $is$ subdominant OP from tunneling measurements. The same but not so evident conclusion was also done in Ref.\,\onlinecite{Kohen} after analysis of the Andreev reflection point-contact spectra.

The maximum strength of the subdominant {\sl s\/}-wave pairing from our measurements is $T_S$$/$$T_C$$\approx$0.24 much higher than $T_S$$/$$T_C$$\approx$0.10 earlier reported \cite{Covington}. Theory \cite{Zhu99} predicts $T_S$$/$$T_C$=0.16.

We have also found that in relativety low magnetic field $\approx$0.04\,T the depth of the minimum and distance between peaks 2$\delta$$_S$ essentially decreases (Fig.\,\ref{Andreev}b). It is reasonable because such magnetic field may effect as a strong depairing factor on the $s$-wave pairing. On the other hand, magnetic field can effect on the Andreev bound states shifting their energies to the higher values (with proper increase of $\delta$$_S$) due to Doppler effect \cite{Covington}. This effect was usually observed earlier. It seems difficult to distinguish these two opposite effects if the minimum at $V$=0 is small (smeared due to roughness of the junction interface, for instance \cite{Kashiwaya}). Nevertheless, looking carefully on the results in Ref.\,\onlinecite{Krupke}, one can find the systematic decrease of {\sl$\delta$$_S$} with the low magnetic field increase too.

\begin{figure}
\begin{center}
\includegraphics[width=9cm,angle=0]{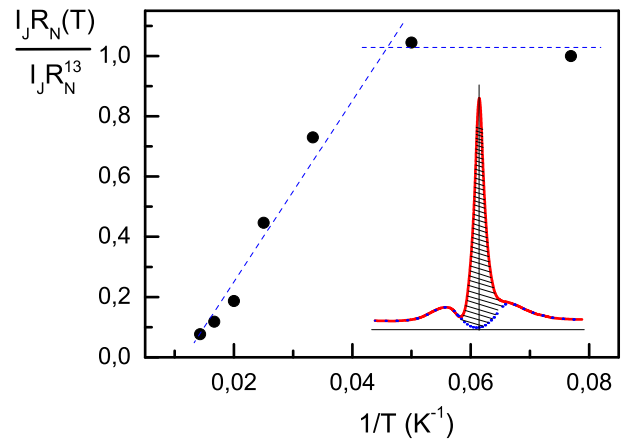}
\end{center}
\caption{Temperature dependence of the Josephson current $I_J$$R_N$ normalized to the value at $T$=13\,K, where $R_N$ is normal state resistance at $V$=100\,mV The current corresponds to a square of the shaded area like shown in a panel.}
\label{TJosephson}
\end{figure}

In addition to this, when the $is$-wave (or $id_{xy}$-wave) subdominant pairing realizes, theory predicts the saturation of the maximum Josephson current at $T$$<$$T_S$ due to the decrease of the density of Andreev bounds states at Fermi level which transfer the Josephson current \cite{Tanaka}. The similar behavior is observed in our experiments as reports in Figs.\,\ref{Ttunneling}\,and\,\ref{TJosephson}. One can see in Fig.\,\ref{Ttunneling} that the width of the narrow peak associated with the Josephson current is almost the same with temperature raise. It says once more that the thermal fluctuation is not a reason of smearing of the Josephson current step. 

It seems reasonable here to characterize the Josephson current by integration of the conductance peak subtracting the possible background (see a panel of Fig.\,\ref{TJosephson}). The temperature dependence of such current is shown in Fig.\,\ref{TJosephson} demonstrating clear saturation at $T$$<$20--30\,K. We have also observed the close to 1/$T$ dependence in large temperature range in agreement with the experimental results for the ZBCP intensity in \mbox{$S$--$I$--$S$} junction in Refs.\,\onlinecite{Alff},\,\onlinecite{Mourachkine} and the theory for the node direction tunneling in the junctions with the same order parameter orientation in both electrodes \cite{Barash}. Such junctions were most probably realized in our experiments.

In summary, the specific form of the tunneling spectrum with the deep minimum around $V$=0, the predicted temperature behavior of the splitted ZBCP and of the Josephson current have been observed in YBCO break-junctions giving the clear evidence for the mixed symmetry {\it d$_{x^{2}-y^{2}}$$\pm$is} of order parameter near the (110) surface in contrast to the {\it d$_{x^{2}-y^{2}}$$\pm$id$_{xy}$} OP. The deduced strength of the subdominant $is$ OP is rather high leading to the transition into the {\it d$_{x^{2}-y^{2}}$$\pm$is} states at 20$<$$T_S$$<$30.

\end{document}